\documentclass[a4paper,12pt]{article}
\usepackage[left=3.17cm,right=3.17cm,top=2.54cm,
headheight=0.5cm,headsep=0.54cm,bottom=2.54cm,footskip=0.79cm
]{geometry}

\usepackage{amsmath,amssymb}

\usepackage{hyperref}

\usepackage{cite}

\usepackage{graphicx}

\begin{document}

\title{Induced entanglement entropy of harmonic oscillators in noncommutative phase space}

\author{Bing-Sheng Lin$^{\,1,\,}$\thanks{Email: sclbs@scut.edu.cn}\,\,,\, Jian Xu$^{\,1}$,\, Tai-Hua Heng$^{\,2}$\\
{\small $^{1\,}$School of Mathematics, South China University of Technology, Guangzhou 510641, China}\\
{\small $^{2\,}$School of Physics and Material Science, Anhui University, Hefei 230601, China}
}

\date{}

\maketitle

\begin{abstract}
We study the entanglement entropy of harmonic oscillators in noncommutative phase space. We propose a new definition of quantum R\'{e}nyi entropy based on Wigner functions in noncommutative phase space. Using the R\'{e}nyi entropy, we calculate the entanglement entropy of the ground state of the $2D$ isotropic harmonic oscillators. We find that for some values of the noncommutative parameters, the harmonic oscillators can be entangled in noncommutative phase space. This is a new entanglement-like effect caused by the noncommutativity of the phase space.
\\

\\

\end{abstract}

\section{Introduction}
In the past decades, there has been much interest in the study of physics in noncommutative
space \cite{Seiberg,Douglas,Chaichian,Ettefaghi,Heng,Kupriyanov,Ghosh,lhc,Calmet,Ghasemkhani,Couch}. The ideas of noncommutative spacetime started in 1947 \cite{Snyder}. In the 1980's,
Connes formulated the mathematically rigorous framework of noncommutative
geometry \cite{Connes}. A noncommutative spacetime also appeared in string theory, namely in the quantization of open string \cite{Seiberg}. The noncommutativity
of spacetime also plays an important role in quantum gravity \cite{Doplicher,Zupnik}. The concept of noncommutative spacetime is also applied in condensed matter physics, such as the integer quantum Hall effect \cite{Polychronakos}. Since the noncommutativity between spatial
and time coordinates may lead to some problems with unitary and causality \cite{Gomis}, usually,
only spatial noncommutativity is considered.
Although in string theory only the coordinate space exhibits a noncommutative structure, some authors have studied models in which a noncommutative
geometry is defined on the whole phase space \cite{Duval,Nair,Banerjee,Zhang,Li,Vakili,Bastos,HMS,Lin,Gnatenko}. Noncommutativity between momenta arises naturally as a consequence of noncommutativity
between positions, as momenta are defined to be the partial derivatives of the action with respect to the position
coordinates.

The noncommutativity between the coordinates of the space may change the properties of the physical systems. For example, in a multipartite system, there maybe some new quantum correlations between the subsystems induced by the noncommutativity of the space, such as quantum entanglement or other types of quantum correlations.
Quantum entanglement is one of the key features of quantum physics, and it has many applications in quantum information, many-body physics and spacetime physics \cite{hhhh,Amico,Laflorencie,Bengtsson,Matsumura}.
Entropy provides a tool that can be used to quantify entanglement. If the overall system is in a pure state, the quantum entropy of the whole system equals zero, and the entropy of one subsystem can be used to measure its degree of entanglement with the other subsystems. This is the so-called entanglement entropy. Some authors have already studied the quantum entanglement and entropy of physical systems in noncommutative space \cite{Jing,Adhikari,Bastos1,Dey,Sabella,Nascimento,Chen}.

One usually uses the von Neumann entropy to analyze the physical systems. The von Neumann entropy is defined by the density operators. Since we consider the physical system in noncommutative phase space (NCPS) in this work, it is convenient to use the Wigner functions to calculate the entropy of the system.
There are some types of quantum entropy defined by the Wigner functions in normal commutative phase space \cite{mf,gse,jjw,zc,skd}. We will use a generalized R\'{e}nyi entropy to analyze the entanglement of the harmonic oscillators in noncommutative phase space. Quantum R\'{e}nyi entropy can be considered as a generalization of von Neumann entropy \cite{renyi}.

This paper is organized as follows. In Sec. \ref{sec2}, we consider the $2D$ harmonic oscillators in noncommutative phase space, and derive the Wigner functions of the system by virtue of deformation quantization method. Using a new definition of quantum R\'{e}nyi entropy in noncommutative phase space, the entanglement entropy of the oscillator system is calculated in Sec. \ref{sec3}. In Sec. \ref{sec4}, we analyze the numerical values of the entanglement entropy of the oscillator system in NCPS.
We find that for some values of the noncommutative parameters, there is entanglement of the harmonic oscillators induced by the noncommutativity of the phase space. Some discussions and conclusions are given in Sec. \ref{sec5}.
The definition of quantum R\'{e}nyi entropy in noncommutative phase space is discussed in Appendix \ref{apa}, and some calculations of the $*-$exponential functions is presented in Appendix \ref{apb}.

\section{Wigner Functions of Harmonic Oscillators in NCPS}\label{sec2}
In this work, we will consider a $4D$ noncommutative phase space in which the coordinate operators $\hat{x}_i$, $\hat{p}_i$ satisfy the following commutation relations \cite{Zhang},
\begin{equation}\label{cr2}
[\hat{x}_i,\,\hat{p}_j]=\mathrm{i}\delta_{ij}\hbar\,,\qquad
[\hat{x}_1,\,\hat{x}_2]=\mathrm{i}\mu\,,\qquad
[\hat{p}_1,\,\hat{p}_2]=\mathrm{i}\nu\,,
\end{equation}
where $i,j=1,2$, and $\mu$, $\nu$ are real parameters. We usually assume that $\mu$, $\nu$ are very small numbers, and $|\mu\nu|\ll \hbar^2$.

Let us consider the simplest $2D$ isotropic harmonic oscillators in the noncommutative phase space, and the Hamiltonian can be
written as
\begin{equation}\label{H0}
H = \frac{1}{2m}p^{2}_{1} + \frac{1}{2m} p^{2}_{2} +
\frac{m\omega^2}{2} x^{2}_{1} + \frac{m\omega^2}{2} x^{2}_{2}~.
\end{equation}

Because of the noncommutativity between the coordinate operators, there are no wave functions such as $\psi(x_1, x_2)$. (The coordinate operators in noncommutative phase space can be transformed into those in normal commutative phase space via the so-called Seiberg-Witten maps, but usually the transformations are not unitary.)
Instead, one can consider the Wigner functions of the system.
By virtue of deformation quantization method \cite{hh}, one can derive the Wigner functions and energy spectra of the system in the noncommutative phase space.
In noncommutative phase space, the Hamiltonian $H$ and the corresponding Wigner functions $W$ satisfy the so-called $\ast-$genvalue equation
\begin{equation}\label{eigen}
    H\ast W =W \ast H=E W,
\end{equation}
where $E$ is the corresponding energy of the physical system, and the Moyal $\ast-$product in NCPS is defined as
\begin{equation}\label{star}
    \ast:=\exp \left\{\frac{\mathrm{i\hbar}}{2}\Big(\overleftarrow{\partial}\!_{x_{i}}\overrightarrow{\partial}\!_{p_{i}}-\overleftarrow{\partial}\!_{p_{i}}\overrightarrow{\partial}\!_{x_{i}}\Big)
    +\frac{\mathrm{i}\mu}{2}\epsilon_{ij}\overleftarrow{\partial}\!_{x_{i}}\overrightarrow{\partial}\!_{x_{j}}
    +\frac{\mathrm{i}\nu}{2}\epsilon_{ij}\overleftarrow{\partial}\!_{p_{i}}\overrightarrow{\partial}\!_{p_{j}}\right\}.
\end{equation}
Here we have used the Einstein summation convention, and $(\epsilon_{ij})$ is the antisymmetric matrix
\[(\epsilon_{ij})=
\begin{pmatrix}
  0 & 1\\
  -1 & 0\\
\end{pmatrix}.
\]
Eq.~(\ref{eigen}) is the so-called $\ast-$genvalue equation, and it corresponds to the time-independent Schr\"odinger equation of wave function. By solving Eq.~(\ref{eigen}), one can obtain the Wigner functions and the energy spectra of $H$.

The Hamiltonian $H$ (\ref{H0}) can be separated into two parts $H_+$ and $H_-$,
\begin{eqnarray}
&&\!\!\!\!\!\!\!\!\!\! H_+=
\frac{1}{2}\left(\frac{p_1}{\sqrt{m}}\cos(c) +\omega\sqrt{m} x_2 \sin(c) \right)^2
+\frac{1}{2}\left( \omega\sqrt{m} x_1\sin(c) -\frac{p_2}{\sqrt{m}}\cos(c)\right)^2,\nonumber\\
&&\!\!\!\!\!\!\!\!\!\! H_-=\frac{1}{2}\left(\frac{p_2}{\sqrt{m}}\sin(c)+ \omega\sqrt{m} x_1\cos(c)\right)^2
+\frac{1}{2}\left( \omega\sqrt{m} x_2 \cos(c) - \frac{p_1}{\sqrt{m}}\sin(c)\right)^2,~~
\end{eqnarray}
where
\begin{equation}\label{asb}
    c=\frac{1}{2}\mathrm{arccot}(\delta),\qquad\delta=\frac{m^2\omega^2\mu-\nu}{2\hbar m \omega}.
\end{equation}
It is easy to verify that $H_+$ being commutative with
$H_-$ under the Moyal bracket
\begin{equation}\label{hhc}
    [H_+~,~H_-]_{\ast}:=H_+\ast H_--H_-\ast H_+=0,
\end{equation}
and their $\ast-$product is equal to their ordinary
product
\begin{equation}\label{hheq}
    H_+\ast H_-=H_+ H_-=H_-\ast H_+~.
\end{equation}

Using the methods in Ref.~\cite{hh}, after some calculations, one can get the Wigner functions and energy spectra of $H_+$ and $H_-$,
\begin{equation}\label{wig}
W^{\pm}_{n}=\frac{(-1)^{n}}{h_\pm\pi}e^{-\frac{2H_\pm}{h_\pm\omega}}
L_{n}\left(\frac{4H_\pm}{h_\pm\omega}\right),
\end{equation}
\begin{equation}
E^{\pm}_{n}=\Big(n+\frac{1}{2}\Big)h_\pm\omega,\qquad (n=0,1,2,...)
\end{equation}
where
\begin{equation}
    h_\pm=\hbar\big(\sqrt{1+\delta^2} \pm \eta\big),\qquad
    \eta=\frac{m^2\omega^2\mu+\nu}{2\hbar m \omega},
\end{equation}
and $L_{n}(x)$ are the Laguerre polynomials.

So the Wigner functions and energy spectrum of $H$ (\ref{H0}) are
\begin{eqnarray}\label{wigh}
    W_{n_1n_2}&=&W^{+}_{n_1}\ast W^{-}_{n_2}=W^{+}_{n_1}W^{-}_{n_2}\nonumber\\
&=&\frac{(-1)^{n_1+n_2}}{\pi^2 h_+ h_-}e^{-\frac{2H_+}{h_+\omega}-\frac{2H_-}{h_-\omega}}
L_{n_1}\left(\frac{4H_+}{h_+\omega}\right)
    L_{n_2}\left(\frac{4H_-}{h_-\omega}\right),
\end{eqnarray}
\begin{eqnarray}
    E_{n_1n_2}&=&E^{+}_{n_1}+E^{-}_{n_2}=\Big(n_1+\frac{1}{2}\Big)h_+\omega+\Big(n_2+\frac{1}{2}\Big)h_-\omega\nonumber\\
    &=&\hbar\omega\Big[(n_1+n_2+1)\sqrt{1+\delta^2}+(n_1-n_2)\eta~\Big].
\end{eqnarray}
It is easy to verify that, the Wigner functions $W_{n_1n_2}$ satisfy the following $*-$orthogonality relations:
\begin{equation}\label{or}
    W_{ij}\ast W_{kl}=\frac{1}{4\pi^2\hbar_+\hbar_-}\delta_{ik}\delta_{jl} W_{ij}
    =\frac{1}{4\pi^2(\hbar^2-\mu\nu)}\delta_{ik}\delta_{jl} W_{ij}~,
\end{equation}
and
\begin{equation}
\int W_{n_1n_2}(x_1,p_1;x_2,p_2)dx_1dx_2dp_1dp_2=1.
\end{equation}

\section{Quantum Entropy and Entanglement of the Harmonic Oscillators in NCPS}\label{sec3}
Now, let us consider the quantum entropy and entanglement of the harmonic oscillators in $4D$ noncommutative phase space.
For a bipartite system, one may use the entanglement entropy, namely, the
entropy of one of its reduced states to measure the entanglement of the system.
We will use the quantum R\'{e}nyi entropy to quantify the entanglement of the harmonic oscillators in the present work. In $4D$ noncommutative phase space, the quantum R\'{e}nyi entropy can be defined by the Wigner functions as follow (see Appendix \ref{apa} for more details),
\begin{equation}\label{sqnc}
S_\alpha(W)=\frac{1}{1-\alpha}\ln\left(\big(4\pi^2(\hbar^2-\mu\nu)\big)^{\alpha-1}\int W^\alpha_{\ast}dx_1dp_1dx_2dp_2\right),
\end{equation}
where $\alpha$ is a positive real parameter, and $W^n_{*}$ is the $n$-th $*-$power of the Wigner function $W$.

In this work, we only consider the cases for $\alpha$ being positive integers. Since there are the orthogonality relations for the pure state Wigner functions such as (\ref{or}), the quantum R\'{e}nyi entropy of pure states equals zero in noncommutative phase space, e.g. $S_\alpha(W_{n_1n_2})=0$. This is just the same as the von Neumann entropy in normal commutative space.

For simplicity, we only consider the entanglement entropy of the ground state of the harmonic oscillators in NCPS.
The corresponding Wigner function is
\begin{equation}
    W_{00}=\frac{1}{\pi^2 h_+ h_-}e^{-\frac{2H_+}{h_+\omega}-\frac{2H_-}{h_-\omega}}~.
\end{equation}
Obviously, this Wigner function is always positive. The Wigner functions of the reduced states are
\begin{eqnarray}\label{w01}
W_{00}^{(1)}(x_1,p_1) &=& \int W_{00}(x_1,p_1;x_2,p_2)dx_2dp_2 \nonumber\\
   &=& \frac{\lambda}{\pi\hbar}e^{-\frac{\sqrt{1+\delta^2} }{\hbar m \omega}\left(\frac{p_1^2}{1+\delta^2-\delta\eta}
   +\frac{m^2 \omega^2x_1^2}{1+\delta^2+\delta\eta}\right)},
\end{eqnarray}
and
\begin{eqnarray}\label{w02}
W_{00}^{(2)}(x_2,p_2)&=&\int W_{00}(x_1,p_1;x_2,p_2)dx_1dp_1\nonumber\\
   &=& \frac{\lambda}{\pi\hbar}e^{-\frac{\sqrt{1+\delta^2} }{\hbar m \omega}\left(\frac{p_2^2}{1+\delta^2-\delta\eta}
   +\frac{m^2 \omega^2x_2^2}{1+\delta^2+\delta\eta}\right)},
\end{eqnarray}
where
\begin{eqnarray}\label{aa}
\lambda&=&\sqrt{\frac{1+\delta^2}{(1+\delta^2)^2-\delta^2\eta^2}}\nonumber\\
&=&\sqrt{\frac{4 \hbar^4 m^2 \omega^2+\hbar^2(m^2 \omega^2\mu -\nu)^2}{4 \hbar^4 m^2 \omega^2+(2 \hbar^2-\mu  \nu)(m^2 \omega^2\mu -\nu)^2}}\,.
\end{eqnarray}
Since the total entropy of the ground state of the $2D$ harmonic oscillators in NCPS is zero,
\begin{equation}
S_\alpha(W_{00})=0,
\end{equation}
the entanglement entropy $E_\alpha(W_{00})$ of the harmonic oscillators is just the entropy of the reduced state (\ref{w01}) or (\ref{w02}).
For a positive integer $\alpha\geqslant 1$, we have
\begin{eqnarray}
E_\alpha(W_{00})&:=& S_\alpha\Big(W_{00}^{(1)}\Big)=S_\alpha\Big(W_{00}^{(2)}\Big)\nonumber\\
&=&\frac{1}{1-\alpha}\ln\left((2\pi \hbar)^{\alpha-1}\int \Big(W_{00}^{(1)}\Big)^\alpha_{*}\,dx_1dp_1\right).
\end{eqnarray}

First, let us consider the cases of positive integers $\alpha\geqslant 2$. Using the $*-$product (\ref{star}), and after some calculations (see Appendix \ref{apb} for more details), one can obtain
\begin{eqnarray}\label{wa}
\lefteqn{\Big(W_{00}^{(1)}\Big)^n_{*}
=\bigg(\frac{\lambda}{\pi\hbar}e^{-\frac{\sqrt{1+\delta^2} }{\hbar m \omega}\left(\frac{p_1^2}{1+\delta^2-\delta\eta}
   +\frac{m^2 \omega^2x_1^2}{1+\delta^2+\delta\eta}\right)}\bigg)_{*}^n}\nonumber\\
&&~~~=\frac{2\lambda^n}{(\pi\hbar)^n\big[(1+\lambda)^n+(1-\lambda)^n\big]}
e^{-\frac{[(1+\lambda)^n-(1-\lambda)^n]\sqrt{1+\delta^2}}{\lambda[(1+\lambda)^n+(1-\lambda)^n]\hbar m \omega}
\left(\frac{p_1^2}{1+\delta^2-\delta\eta}
   +\frac{m^2 \omega^2x_1^2}{1+\delta^2+\delta\eta}\right)}.
\end{eqnarray}
Integrating the above function (\ref{wa}), we will get the following result:
\begin{eqnarray}\label{sq}
E_\alpha(W_{00})
&=&\frac{1}{1-\alpha}\ln\left(\frac{(2\lambda)^{\alpha}}{(1+\lambda)^\alpha-(1-\lambda)^\alpha}\right)\nonumber\\
&=&\frac{1}{\alpha-1}\ln\big[(1+\lambda)^\alpha-(1-\lambda)^\alpha\big]-\frac{\alpha}{\alpha-1}\ln(2\lambda).
\end{eqnarray}

For $\alpha=1$, the expression of the quantum R\'{e}nyi entropy of the reduced state will become (see Appendix \ref{apa} for more details)
\begin{eqnarray}\label{s11}
S_1\Big(W_{00}^{(1)}\Big)&=&-\int W_{00}^{(1)}*\ln_{*}\!\Big(2\pi\hbar\, W_{00}^{(1)}\Big)dx_1dp_1\nonumber\\
&=&-\int W_{00}^{(1)}\ln_{*}\!\Big(2\pi\hbar\, W_{00}^{(1)}\Big)dx_1dp_1.
\end{eqnarray}
This expression just corresponds to the von Neumann entropy in phase space.

The reduced state Wigner function $W_{00}^{(1)}$ (\ref{w01}) can be rewritten as (see Appendix \ref{apb} for more details)
\begin{equation}
W_{00}^{(1)}= \frac{\lambda}{\pi\hbar \sqrt{1-\lambda^2}}
   \,\mathrm{exp}_*\!\left[\frac{\sqrt{1+\delta^2} }{2\lambda\hbar m \omega}\left(\frac{p_1^2}{1{+}\delta^2{-}\delta\eta}
   +\frac{m^2 \omega^2x_1^2}{1{+}\delta^2{+}\delta\eta}\right)\ln\left(\frac{1-\lambda}{1+\lambda}\right)\right].
\end{equation}
So we have
\begin{eqnarray}
\lefteqn{\ln_{*}\!\Big(2\pi\hbar\, W_{00}^{(1)}\Big)}\nonumber\\
&&~~~~=\ln_{*}\!\left(\frac{2\lambda}{\sqrt{1-\lambda^2}}\,\mathrm{exp}_*\!\left[\frac{\sqrt{1+\delta^2} }{2\lambda\hbar m \omega}\left(\frac{p_1^2}{1{+}\delta^2{-}\delta\eta}
   +\frac{m^2 \omega^2x_1^2}{1{+}\delta^2{+}\delta\eta}\right)\ln\!\left(\frac{1-\lambda}{1+\lambda}\right)\right]\right)\nonumber\\
&&~~~~=\ln\!\left(\frac{2\lambda}{\sqrt{1-\lambda^2}}\right)+\frac{\sqrt{1+\delta^2} }{2\lambda\hbar m \omega}\left(\frac{p_1^2}{1{+}\delta^2{-}\delta\eta}
   +\frac{m^2 \omega^2x_1^2}{1{+}\delta^2{+}\delta\eta}\right)\ln\!\left(\frac{1-\lambda}{1+\lambda}\right).
\end{eqnarray}
After some straightforward calculations, one can obtain the entanglement entropy $E_1(W_{00})$,
\begin{eqnarray}\label{s12}
E_1(W_{00})&:=& S_1\Big(W_{00}^{(1)}\Big)\nonumber\\
&=&-\ln\left(\frac{2\lambda}{\sqrt{1-\lambda^2}}\right)
-\frac{1}{2\lambda}\ln\left(\frac{1-\lambda}{1+\lambda}\right)\nonumber\\
&=&\frac{1}{2\lambda}\big[(1+\lambda)\ln(1+\lambda)-(1-\lambda)\ln(1-\lambda)\big]-\ln(2\lambda).
\end{eqnarray}
This is just the von Neumann entropy of the reduced state $W_{00}^{(1)}$.
It is easy to verify that the expression (\ref{sq}) will reduce to the above expression (\ref{s12}) when $\alpha\to 1$. So in this sense, one can regard the formula (\ref{sq}) as the general expression of the entanglement entropy of the oscillator system for all positive integers $\alpha=1,2,3,...$

\section{Some Numerical Results}\label{sec4}
Since $\eta^2-\delta^2=\mu\nu/\hbar^2$, if we assume $|\mu\nu|\ll \hbar^2$, then $-1<\delta^2-\eta^2<1$.
From the expression (\ref{aa}), it is easy to see that
\begin{equation}
0.577=\frac{\sqrt{3}}{3}<\lambda\leqslant 1.
\end{equation}
When $\delta=0$ or $1+\delta^2-\eta^2=0$, namely, $\mu=\nu=0$ or $\nu/\mu = m^2 \omega^2$ or $\mu\nu=\hbar^2$,
there is $\lambda=1$.
When $\delta^2-\eta^2\to 1$ and $\delta^2\to \infty$, namely, $\mu\nu\to -\hbar^2$ and $|m^2\omega^2\mu-\nu|\to \infty$,
there is $\lambda\to \sqrt{3}/3$.

From Eq.~(\ref{sq}), one can obtain
\begin{equation}
\frac{\partial}{\partial\lambda}E_\alpha(W_{00})< 0,\qquad \frac{\partial}{\partial\alpha}E_\alpha(W_{00})\leqslant 0,
\end{equation}
and the equality holds if and only if $\lambda=1$.
So from the expression (\ref{sq}) and (\ref{s12}), for $\lambda\in (\sqrt{3}/3,1]$ and $\alpha=1,2,3,...$, we have
\begin{equation}
E_\alpha(W_{00})(\lambda)\geqslant E_{\alpha+1}(W_{00})(\lambda),
\end{equation}
and
\begin{equation}
0\leqslant E_\alpha(W_{00})<\frac{\sqrt{3}}{2}\ln\big(2+\sqrt{3}\big)-\frac{1}{2}\ln2=0.794,
\end{equation}
and the equalities hold if and only if $\lambda=1$.
So the entanglement entropy $E_\alpha(W_{00})$ is always nonnegative.

\begin{figure}[!ht]
\centering
\includegraphics[width=0.60\textwidth]{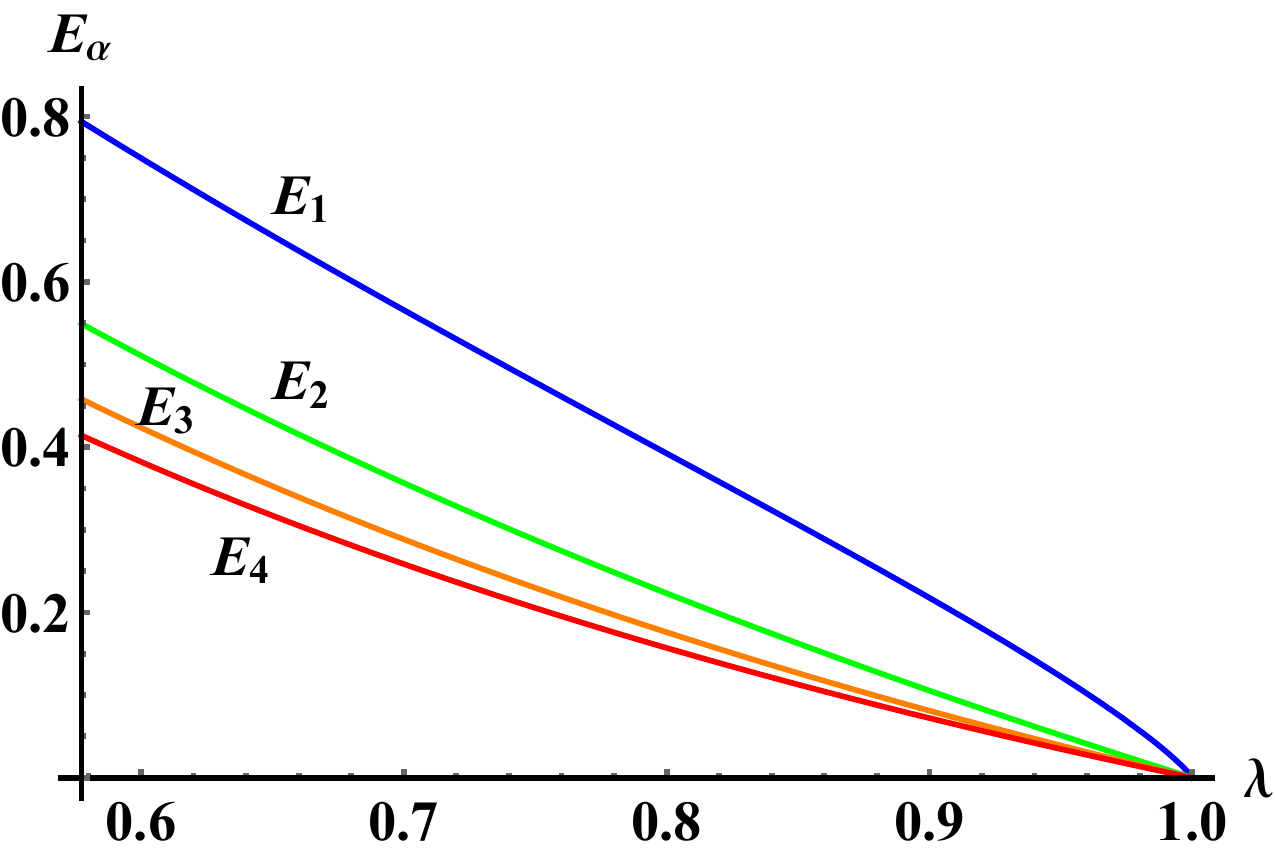}
\caption{\label{fig1}The entanglement entropy $E_\alpha(W_{00})$, with respect to the variable $\lambda$. $E_1$ is the von Neumann entropy of the reduced states.}
\end{figure}

Fig.~\ref{fig1} shows the entanglement entropy $E_\alpha(W_{00})$ with respect to the variable $\lambda$, and $\alpha=1,2,3,4$.
$E_1$ is just the von Neumann entropy of the reduced states. Obviously, we have $E_1\geqslant E_2\geqslant E_3\geqslant E_4$, and the equalities hold if and only if $\lambda=1$. It is known that the entropy measures the amount of information about the system. So this means that, in general, the quantum R\'{e}nyi entropy $S_\alpha(W)$ with smaller number $\alpha$ can give us more information about the physical systems.

When $\lambda=1$, namely, $\mu=\nu=0$ or $\nu/\mu = m^2 \omega^2$ or $\mu\nu=\hbar^2$, the entanglement entropy of the harmonic oscillators reaches its minimum $E_\alpha^{\text{min}}(W_{00})=0$.
This means that there is no entanglement in the oscillator system. $\mu=\nu=0$ is just the case in normal commutative space.
$\mu\nu=\hbar^2$ will cause some singularity, and we usually assume $|\mu\nu|\ll \hbar^2$.
For the case $\nu/\mu = m^2 \omega^2$, there is also no entanglement in the system, while there is the noncommutativity in the phase space. But in our opinion, the parameters $\mu$ and $\nu$ reflect the intrinsic noncommutativity
between positions and momenta, respectively (just like the Planck constant encodes the noncommutativity of position and momentum), which should be independent on the parameters of concrete physical models.

In other cases, the entanglement entropy of the system is always positive, $E_\alpha(W_{00})>0$.
This means that the subsystems are mixed states while the whole system is in a pure state. So there is entanglement in the harmonic oscillators in noncommutative phase space, while it vanishes in normal commutative phase space. This is an entanglement-like effect caused by the noncommutativity of the phase space.
To our knowledge, this result has not been reported in the literatures.

When $\lambda\to \sqrt{3}/3$, namely, $\mu\nu\to -\hbar^2$ and $|m^2\omega^2\mu-\nu|\to \infty$, the entanglement entropy approaches its maximum.

\begin{figure}[!ht]
\centering
\includegraphics[width=0.60\textwidth]{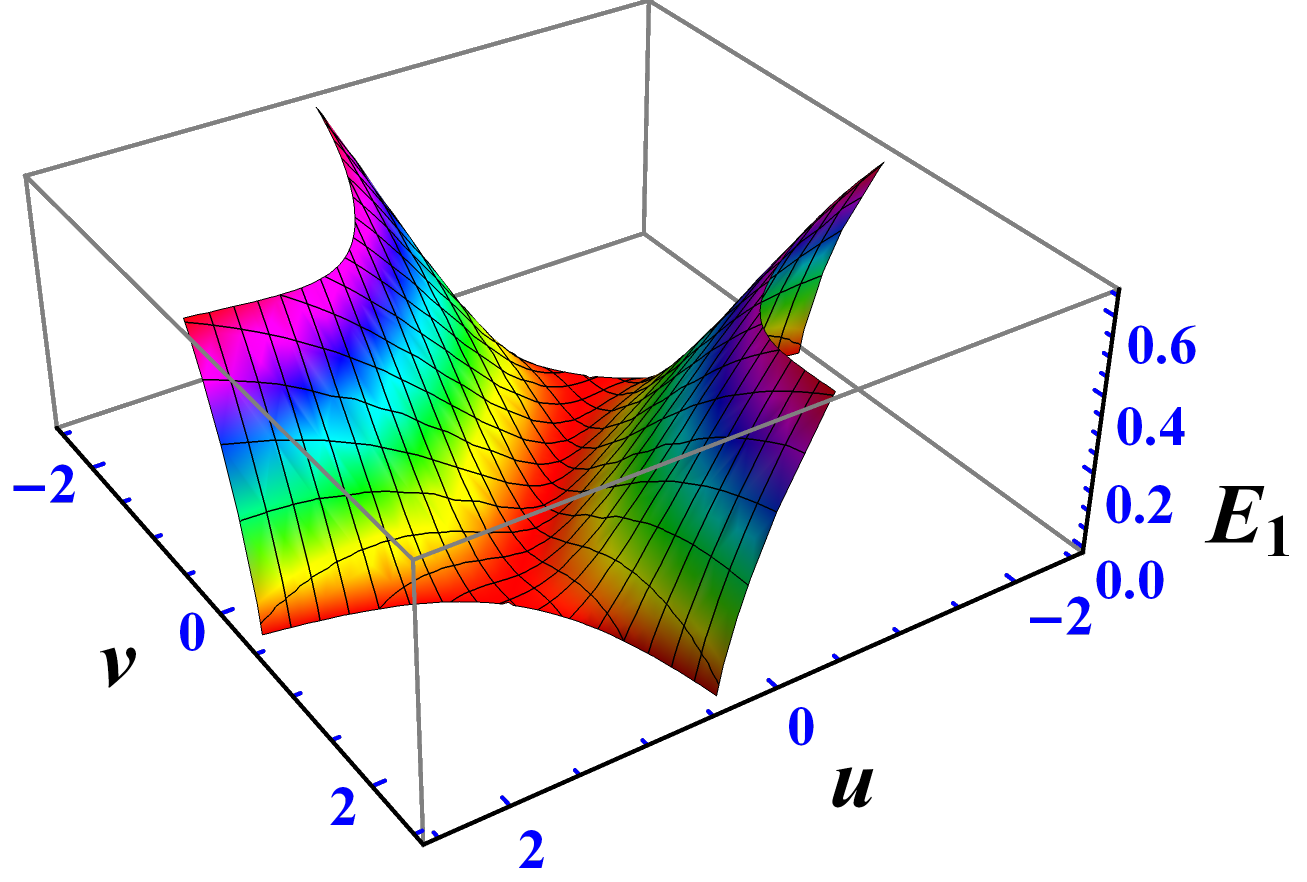}
\caption{\label{fig2}The entanglement entropy $E_1(W_{00})$, with respect to the variables $u$ and $v$.}
\end{figure}

From the expression (\ref{aa}), we can rewrite $\lambda$ as
\begin{equation}
\lambda=\sqrt{\frac{4 + (u -v)^2}{4 +(2 -uv)(u -v)^2}}\,,
\end{equation}
where
\begin{equation}
u=\frac{m \omega \mu}{\hbar},\qquad v=\frac{\nu}{\hbar m \omega}~.
\end{equation}
Fig.~\ref{fig2} shows the entanglement entropy $E_1(W_{00})$ with respect to the variables $u$ and $v$ (here we assume $-1<uv<1$).

\begin{figure}[!ht]
\centering
\includegraphics[width=0.60\textwidth]{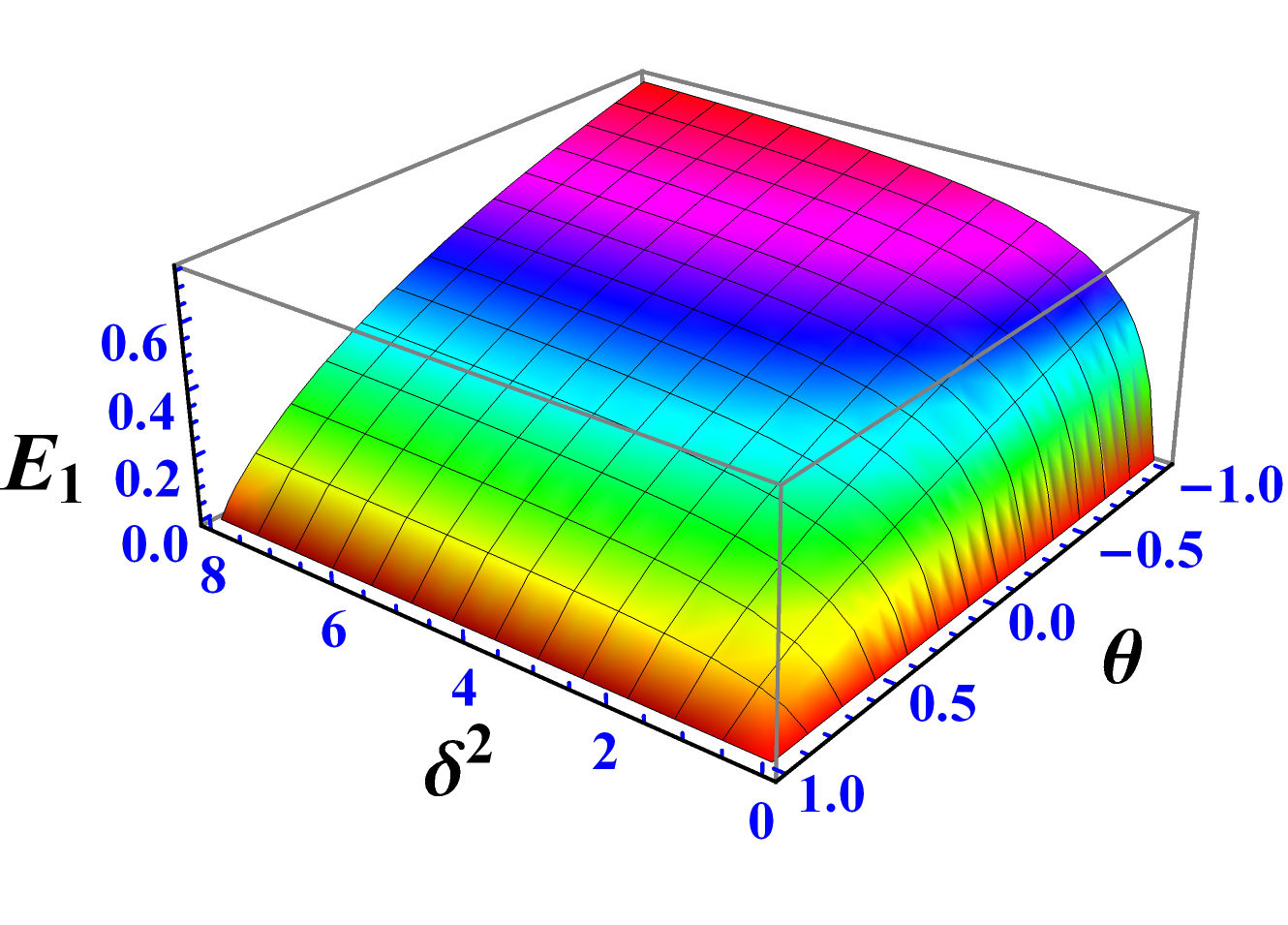}
\caption{\label{fig3}The entanglement entropy $E_1(W_{00})$, with respect to the variables $\delta^2$ and $\theta$.}
\end{figure}

Denote $\theta=\mu\nu/\hbar^2=\eta^2-\delta^2$, then $-1<\theta<1$. We can also rewrite $\lambda$ as
\begin{equation}
\lambda=\sqrt{\frac{1+\delta^2}{1+(2-\theta)\delta^2}}~.
\end{equation}
Fig.~\ref{fig3} shows $E_1(W_{00})$ with respect to the variables $\delta^2$ and $\theta$.
One can also plot the figures of $E_\alpha(W_{00})$ with other parameters $\alpha$, which will be very similar to Figs.~\ref{fig2} and \ref{fig3}.

One can also calculate the entanglement entropy for the excited states of the harmonic oscillators in the noncommutative phase space, but usually the results are much more complex.

\begin{figure}[!ht]
\centering
\includegraphics[width=0.60\textwidth]{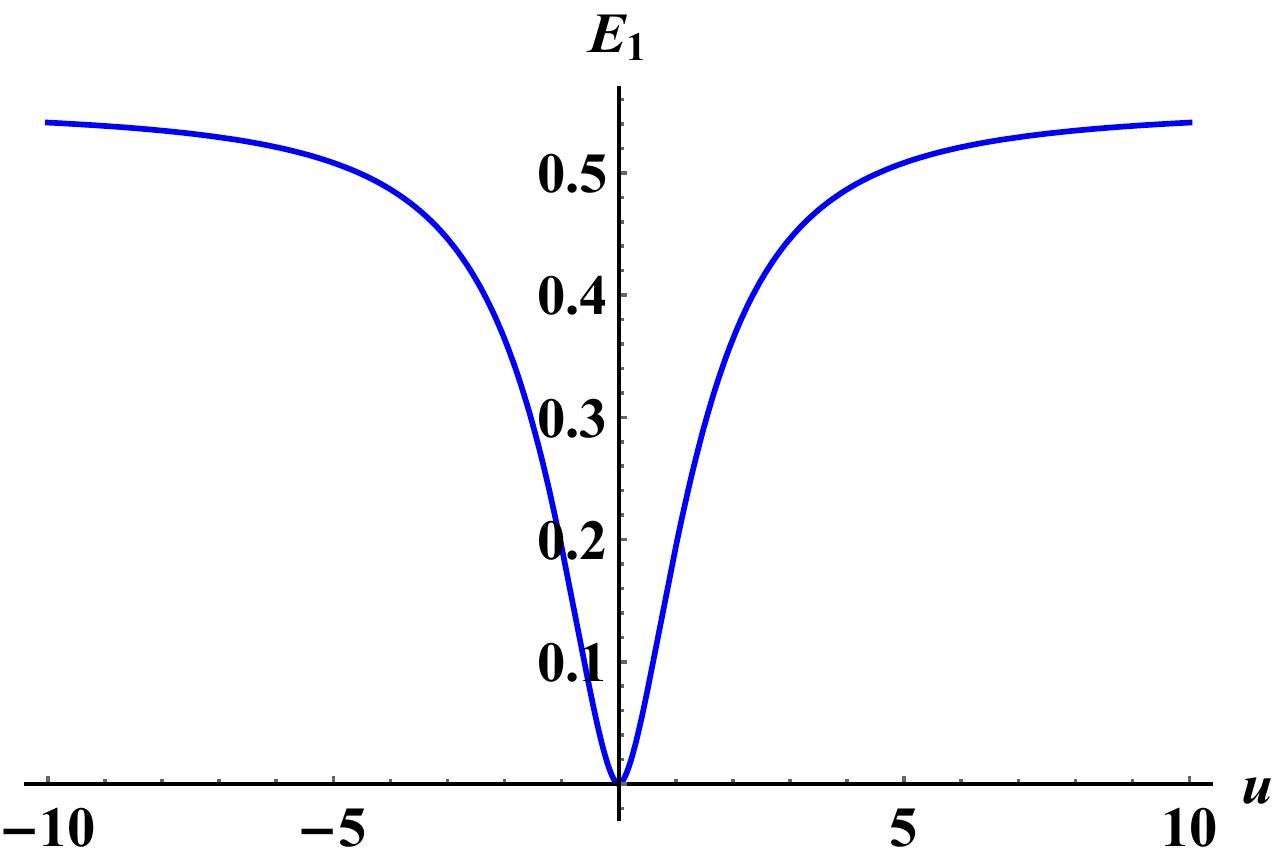}
\caption{\label{fig4}In the case $\mu\neq 0$ and $\nu=0$, the entanglement entropy $E_1(W_{00})$ with respect to the variables $u$.}
\end{figure}

Furthermore, one can also consider the special case $\mu\neq 0$ and $\nu=0$, and the commutation relations of the coordinate operators are
\begin{equation}
[\hat{x}_i,\,\hat{p}_j]=\mathrm{i}\delta_{ij}\hbar\,,\qquad
[\hat{x}_1,\,\hat{x}_2]=\mathrm{i}\mu\,,\qquad
[\hat{p}_1,\,\hat{p}_2]=0\,.
\end{equation}
This is one of the cases studied most frequently in the literatures \cite{Seiberg}.
In this case, we have
\begin{equation}
\delta=\eta=\frac{1}{2}u=\frac{m \omega \mu}{2\hbar}~,\qquad \lambda=\sqrt{\frac{1+\delta^2}{1+2\delta^2}}
=\sqrt{\frac{4 + u^2}{4+2u^2}}~,
\end{equation}
and
\begin{equation}
0.707=\frac{\sqrt{2}}{2}<\lambda\leqslant 1.
\end{equation}
The entanglement entropy of the oscillator system can be expressed as
\begin{eqnarray}
E_\alpha(W_{00})
&=&\frac{1}{\alpha-1}\ln\left[\Big(\sqrt{4+2u^2}+\sqrt{4+ u^2}\Big)^\alpha-\Big(\sqrt{4+2 u^2}-\sqrt{4+u^2}\Big)^\alpha\right]\nonumber\\
&&~~~-\frac{\alpha}{\alpha-1}\ln\Big(2\sqrt{4+u^2}\Big).
\end{eqnarray}
It is easy to see that in this case we have
\begin{equation}
0\leqslant E_\alpha(W_{00})<\sqrt{2}\ln\left(1+\sqrt{2}\right)-\ln(2)=0.553.
\end{equation}
The entanglement entropy $E_1$ of the
harmonic oscillators in this case is plotted in Fig.~\ref{fig4}.
Obviously, the entanglement entropy $E_1$ becomes larger as the absolute value of $u$ increases. This means that the entanglement of the harmonic oscillators increases with the increase of the noncommutativity of the position space.

For the case $\mu=0$ and $\nu\neq 0$, one will obtain similar results.

\section{Discussions and Conclusions}\label{sec5}
In this paper, we study the entanglement entropy of the $2D$ isotropic harmonic oscillators in noncommutative phase space. By virtue of the deformation quantization method, we obtain the Wigner functions of the harmonic oscillators in NCPS. We propose a new definition of the quantum R\'{e}nyi entropy based on Wigner functions in noncommutative phase space.
Using the R\'{e}nyi entropy, we calculate the quantum entropy of the ground state of the harmonic oscillators. We find that in noncommutative phase space, the entropy of the whole system equals zero, but the entropy of the reduced states can be nonzero for some values of the noncommutative parameters. This means that the whole system is in a pure state while the reduced states are mixed states. So the $2D$ isotropic harmonic oscillators can be entangled in noncommutative phase space. This is a new entanglement-like effect caused by the noncommutativity of the phase space. On the other hand, for some special values of the noncommutative parameters, the entanglement can vanish. So one may determine the values of the noncommutative parameters by measuring the entanglement of the oscillator system. To our knowledge, these results have not been reported in the literatures.

One can also use this method to calculate the entanglement entropy for the excited states of the harmonic oscillators in the noncommutative phase space. But usually, the calculations will be very cumbersome, and the results will be much more complex.

In fact, one can also transform the coordinate operators in noncommutative phase space into those in normal commutative phase space via the so-called Seiberg-Witten maps, and then calculate the quantum entropy of the physical systems. But usually the transformations are not unitary. Quantum entropy is just invariant under unitary transformations. Here we use the Wigner functions directly to calculate the quantum entropy of the physical systems in noncommutative phase space, and do not have to rely on these kinds of non-unitary transformations.

Our results and methods can be generalized to the cases of other physical systems in higher-dimensional noncommutative phase space. Since the quantum entanglement has many applications in quantum information and other physical areas, we hope that our results can help to study the physical properties of noncommutative phase space. One can also use these results to test the entanglement of the harmonic oscillators and then examine the noncommutativity of the phase space by designing some experiments.

\section*{Acknowledgements}
This work is supported by the National Natural Science Foundation of China (Grants No. 11405060 and No. 11571119) and the Fundamental Research Funds for the Central Universities (Grant No. 2019MS109).

\appendix

\section{Quantum R\'{e}nyi Entropy in Commutative and Noncommutative Phase Space}\label{apa}
The quantum R\'{e}nyi entropy is a generalization of von Neumann entropy \cite{renyi}, it can be defined as
\begin{equation}
\mathcal{S}_\alpha(\rho)=\frac{1}{1-\alpha}\ln\big(\mathrm{Tr}(\rho^\alpha)\big),
\end{equation}
where $\alpha$ is some real positive parameter, and $\rho$ is the density operator. In the limit for $\alpha\to 1$, the quantum R\'{e}nyi entropy is just the von Neumann entropy,
\begin{equation}
\mathcal{S}_1(\rho)=-\mathrm{Tr}\big(\rho\ln(\rho)\big).
\end{equation}
In normal commutative phase space, the quantum R\'{e}nyi entropy can be defined as \cite{zc},
\begin{equation}\label{sqc}
\mathcal{S}_\alpha(\mathcal{W})=\frac{1}{1-\alpha}\ln\left(\big((2\pi\hbar)^d\big)^{\alpha-1}\int \mathcal{W}^\alpha_{\star}(\boldsymbol{y},\boldsymbol{q})d\boldsymbol{y}d\boldsymbol{q}\right),
\end{equation}
where $d$ is the number of degrees of freedom of the system under consideration, and ``$2\pi\hbar$'' is from the size of the minimal phase space cell $\Delta y_i\Delta q_i$ \cite{mf}. $\mathcal{W}(\boldsymbol{y},\boldsymbol{q})$ is the Wigner function of the system in commutative phase space.
In normal commutative phase space, the coordinate operators $\hat{y}_i$, $\hat{q}_i$ satisfy the standard commutation relations
\begin{equation}\label{cr1}
[\hat{y}_i,\,\hat{q}_j]=\mathrm{i}\delta_{ij}\hbar\,,\qquad
[\hat{y}_i,\,\hat{y}_j]=0\,,\qquad
[\hat{q}_i,\,\hat{q}_j]=0\,,
\end{equation}
and the normal Moyal star product ``$\star$'' is defined as
\begin{equation}
    \star:=\exp \left\{\sum_{i}\frac{\mathrm{i\hbar}}{2}\Big(\overleftarrow{\partial}\!_{y_{i}}\overrightarrow{\partial}\!_{q_{i}}
    -\overleftarrow{\partial}\!_{q_{i}}\overrightarrow{\partial}\!_{y_{i}}
    \Big)\right\}.
\end{equation}
$\mathcal{W}^n_{\star}$ is the $n$-th $\star-$power of the Wigner function $\mathcal{W}$,
\begin{equation}
\mathcal{W}^n_{\star}=\underbrace{\mathcal{W}\star \mathcal{W}\star...\star \mathcal{W}}_n.
\end{equation}

For the pure states, the corresponding Wigner functions satisfy the orthogonality relations $(2\pi\hbar)^d\,\mathcal{W}\star \mathcal{W}=\mathcal{W}$ \cite{zfc}. So in commutative phase space, we have zero R\'{e}nyi entropy for the pure states $\mathcal{W}$, $\mathcal{S}_\alpha(\mathcal{W})=0$. This is just the same as the von Neumann entropy in normal quantum mechanics.

When $\alpha\to1$, the entropy $\mathcal{S}_\alpha(\mathcal{W})$ defined above will reduce to the following,
\begin{equation}\label{s1}
\mathcal{S}_1(\mathcal{W})=-\int \mathcal{W}\star\ln_{\star}\!\big((2\pi\hbar)^d \mathcal{W}\big)d\boldsymbol{y}d\boldsymbol{q},
\end{equation}
where the $\star-$logarithm is
\begin{equation}
\ln_\star(f):=-\sum_{n=1}^{\infty}\frac{(1-f)^n_{\star}}{n}.
\end{equation}
The expression (\ref{s1}) corresponds to the von Neumann entropy in phase space, and it has already been studied in Ref.~\cite{zc}.

Now let us consider the quantum R\'{e}nyi entropy in $4D$ noncommutative phase space.
In noncommutative phase space, the coordinate operators $\hat{x}_{i}$ and $\hat{p}_{i}$ satisfy the extended commutation relations (\ref{cr2}).
Consider some transformation between the coordinates $(y_1,y_2,q_1,q_2)$ of commutative phase space and the coordinates $(x_1,x_2,p_1,p_2)$ of noncommutative phase space,
\begin{equation}\label{map}
\begin{pmatrix} x_{1} \\ x_{2} \\
p_{1} \\ p_{2} \end{pmatrix} = M \begin{pmatrix} y_{1} \\ y_{2} \\ q_{1} \\
q_{2} \end{pmatrix},
\end{equation}
where $M$ is the corresponding transformation matrix.
Using the commutation relations (\ref{cr2}) and (\ref{cr1}), one can derive the following relations:
\begin{equation}\label{eq3}
\begin{pmatrix}
 0 & \mathrm{i}\mu & \mathrm{i}\hbar & 0 \\
 -\mathrm{i}\mu & 0 & 0 & \mathrm{i}\hbar \\
 -\mathrm{i}\hbar & 0 & 0 & \mathrm{i}\nu \\
 0 & -\mathrm{i}\hbar & -\mathrm{i}\nu & 0
\end{pmatrix}=M\begin{pmatrix}
 0 & 0 & \mathrm{i}\hbar & 0 \\
 0 & 0 & 0 & \mathrm{i}\hbar \\
 -\mathrm{i}\hbar & 0 & 0 & 0 \\
 0 & -\mathrm{i}\hbar & 0 & 0
\end{pmatrix}M^T.
\end{equation}
It is easy to derive the determinant of $M$,
\begin{equation}
|M|=1-\frac{\mu\nu}{\hbar^2}.
\end{equation}

So in the $4D$ noncommutative phase space, the size of the minimal phase space cell can be considered as
\begin{eqnarray}
\Delta x_1\Delta p_1\Delta x_2\Delta p_2&=&|M|\Delta y_1\Delta q_1\Delta y_2\Delta q_2\nonumber\\
&=&\left(1-\frac{\mu\nu}{\hbar^2}\right)(2\pi \hbar)^2=4\pi^2(\hbar^2-\mu\nu).
\end{eqnarray}
Similar to the definition (\ref{sqc}), the quantum R\'{e}nyi entropy in $4D$ noncommutative phase space can be defined as
\begin{eqnarray}
S_\alpha(W)&=&\frac{1}{1-\alpha}\ln\left(\big(|M|4\pi^2\hbar^2\big)^{\alpha-1}\int W^\alpha_{\ast}dx_1dp_1dx_2dp_2\right)\nonumber\\
&=&\frac{1}{1-\alpha}\ln\left(\big(4\pi^2(\hbar^2-\mu\nu)\big)^{\alpha-1}\int W^\alpha_{\ast}dx_1dp_1dx_2dp_2\right).
\end{eqnarray}
This result can be generalized to the cases in higher-dimensional noncommutative phase space.

\section{The $*-$exponential Functions}\label{apb}
Define the $\ast-$exponential function as follows \cite{zfc},
\begin{equation}\label{expd}
    \mathrm{exp}_*(Ht):=\sum_{n=0}^{\infty}\frac{t^{n}}{n!}H_\ast^{n},
\end{equation}
where $H$ is a Hamiltonian, and $t$ is some parameter.
Let us consider the following special case:
\begin{equation}
H=(a_ix_i+b_ip_i)^2+(c_ix_i+d_ip_i)^2,
\end{equation}
here $i=1,2$, and we have used the Einstein summation convention.
Using the $*-$product (\ref{star}), we have
\begin{eqnarray}\label{ph}
\lefteqn{\frac{d}{dt}\mathrm{exp}_*(Ht)=H\ast\mathrm{exp}_*(Ht)}\nonumber\\
&&~~~~=\left[\left(a_i\Big(x_{i}\!+\!\frac{\mathrm{i}\hbar}{2}\overrightarrow{\partial}\!_{p_i}\!+\!\frac{\mathrm{i}\mu}{2}\epsilon_{ij}\!\overrightarrow{\partial}\!_{x_j}\Big)
    \!+\!b_i\Big(p_{i}\!-\!\frac{\mathrm{i}\hbar}{2}\overrightarrow{\partial}\!_{x_i}\!+\!\frac{\mathrm{i}\nu}{2}\epsilon_{ij}\!\overrightarrow{\partial}\!_{p_j}\Big)\right)^2\right.\nonumber\\
    &&~~~~~~\left.+\!\left(c_i\Big(x_{i}\!+\!\frac{\mathrm{i}\hbar}{2}\overrightarrow{\partial}\!_{p_i}\!+\!\frac{\mathrm{i}\mu}{2}\epsilon_{ij}\!\overrightarrow{\partial}\!_{x_j}\Big)
    \!+\!d_i\Big(p_{i}\!-\!\frac{\mathrm{i}\hbar}{2}\overrightarrow{\partial}\!_{x_i}\!+\!\frac{\mathrm{i}\nu}{2}\epsilon_{ij}\!\overrightarrow{\partial}\!_{p_j}\Big)\right)^2\right]\mathrm{exp}_*(Ht)\nonumber\\
&&~~~~=\Big(H-k^2\partial_{\!_H}-k^2H\partial_{\!_H}^{\,2}\Big)\mathrm{exp}_*(Ht),
\end{eqnarray}
where
\begin{eqnarray}\label{kk}
k&=&(a_1d_1+a_2d_2-b_1c_1-b_2c_2)\hbar+(a_1c_2-a_2c_1)\mu+(b_1d_2-b_2d_1)\nu\nonumber\\
&=&(\boldsymbol{a}\cdot\boldsymbol{d}-\boldsymbol{b}\cdot\boldsymbol{c})\,\hbar
+(\boldsymbol{a}\wedge\boldsymbol{c})\,\mu+(\boldsymbol{b}\wedge\boldsymbol{d})\,\nu,
\end{eqnarray}
and $\boldsymbol{a}=\{a_1,a_2\}$,
$\boldsymbol{b}=\{b_1,b_2\}$,
$\boldsymbol{c}=\{c_1,c_2\}$ and
$\boldsymbol{d}=\{d_1,d_2\}$.
The solution of the above differential equation can be expressed as
\begin{equation}\label{sexp}
   \mathrm{exp}_*(Ht)
   =\frac{1}{\cosh(kt)}\exp\left(\frac{H}{k}\tanh(kt)\right).
\end{equation}
The value of the parameter $t$ can be chosen as
\begin{equation}
t=\frac{\tanh^{-1}(k)}{k},\quad \mathrm{or} \quad\frac{\tanh(k t)}{k}=1,
\end{equation}
and we have
\begin{eqnarray}
\exp(H)
&=&\cosh\big(\tanh^{-1}(k)\big)\mathrm{exp}_*\left(\frac{H}{k}\tanh^{-1}(k)\right)\nonumber\\
&=&\frac{1}{\sqrt{1-k^2}}\,\mathrm{exp}_*\!\left[\frac{H}{2k}
\ln\left(\frac{1+k}{1-k}\right)\right].
\end{eqnarray}
Furthermore, we also have
\begin{eqnarray}
\big(\exp(H)\big)_*^n&:=&\exp(H)*\exp(H)*...*\exp(H)\nonumber\\
&=&\left(\frac{1}{\sqrt{1-k^2}}\,\mathrm{exp}_*\!\left[\frac{H}{2k}
\ln\left(\frac{1+k}{1-k}\right)\right]\right)_*^n\nonumber\\
&=&(1-k^2)^{-\frac{n}{2}}\,\mathrm{exp}_*\!\left[\frac{nH}{2k}
\ln\left(\frac{1+k}{1-k}\right)\right]\nonumber\\
&=&\frac{2}{(1+k)^n+(1-k)^n}\,\mathrm{exp}\!\left(H
\frac{(1+k)^n-(1-k)^n}{k[(1+k)^n+(1-k)^n]}\right).
\end{eqnarray}

Let us consider the simplest case $H=a x_1^2+b p_1^2$. There is $k=\hbar\sqrt{ab}$, and
\begin{equation}
\exp(a x_1^2+b p_1^2)
=\frac{1}{\sqrt{1-\hbar^2 ab}}\,\mathrm{exp}_*\!\left[\frac{a x_1^2+b p_1^2}{2\hbar\sqrt{ab}}
\ln\left(\frac{1+\hbar\sqrt{ab}}{1-\hbar\sqrt{ab}}\right)\right].
\end{equation}
So for the reduced state Wigner function $W_{00}^{(1)}(x_1,p_1)$ (\ref{w01}), we have
\begin{eqnarray}
\lefteqn{W_{00}^{(1)}(x_1,p_1) = \frac{\lambda}{\pi\hbar}e^{-\frac{\sqrt{1+\delta^2} }{\hbar m \omega}\left(\frac{p_1^2}{1+\delta^2-\delta\eta}
   +\frac{m^2 \omega^2x_1^2}{1+\delta^2+\delta\eta}\right)} }\nonumber\\
   &&= \frac{\lambda}{\pi\hbar \sqrt{1-\lambda^2}}
   \,\mathrm{exp}_*\!\left[\frac{\sqrt{1+\delta^2} }{2\lambda\hbar m \omega}\left(\frac{p_1^2}{1+\delta^2-\delta\eta}
   +\frac{m^2 \omega^2x_1^2}{1+\delta^2+\delta\eta}\right)\ln\left(\frac{1-\lambda}{1+\lambda}\right)\right]\!\!,~~~~
\end{eqnarray}
and
\begin{eqnarray}
\Big(W_{00}^{(1)}\Big)^n_{*}
&=&\bigg(\frac{\lambda}{\pi\hbar}e^{-\frac{\sqrt{1+\delta^2} }{\hbar m \omega}\left(\frac{p_1^2}{1+\delta^2-\delta\eta}
   +\frac{m^2 \omega^2x_1^2}{1+\delta^2+\delta\eta}\right)}\bigg)_{*}^n\nonumber\\
&=&\frac{\lambda^n}{(\pi\hbar)^n}\left(\mathrm{exp}\!\left[-\frac{\sqrt{1+\delta^2} }{\hbar m \omega}\left(\frac{p_1^2}{1+\delta^2-\delta\eta}
   +\frac{m^2 \omega^2x_1^2}{1+\delta^2+\delta\eta}\right)\right]\right)_{*}^n\\
&=&\frac{2\lambda^n}{(\pi\hbar)^n\big[(1+\lambda)^n+(1-\lambda)^n\big]}
e^{-\frac{[(1+\lambda)^n-(1-\lambda)^n]\sqrt{1+\delta^2}}{\lambda[(1+\lambda)^n+(1-\lambda)^n]\hbar m \omega}
\left(\frac{p_1^2}{1+\delta^2-\delta\eta}
   +\frac{m^2 \omega^2x_1^2}{1+\delta^2+\delta\eta}\right)}.\nonumber
\end{eqnarray}

\end{document}